# Assessing the Circadian Rhythm of Cats Living in a Group using Accelerometers

*Alix Enault[1,2], Alia Chebly[3], Leslie Moinet[3], Thierry Bedossa[4], Sarah Jeannin[2] & Thierry Legou[1]*

[1]*Laboratoire Parole et Langage, 5 avenue Pasteur, 13100 Aix-en-Provence, France*
thierry.legou@lpl-aix.fr, enaultalix@gmail.com
[2]*LECD Université Paris Nanterre, 200 avenue de la République 92000 Nanterre, France*
sarah.jeannin@gmail.com
[3]*Blackfoot, 18 rue Pasteur, 94270 Le Kremlin-Bicêtre, France* aliachebly@gmail.com,
lesliemoinet@gmail.com
[4]*Clinique du Pont de Neuilly - Association Agir pour la Vie Animale, 40 Le Quesnoy, 76220 Cuy-Saint-Fiacre, France* tbedossa@yahoo.fr

## ABSTRACT

This study explores the biological rhythms of domestic cats. Twelve cats from the AVA shelter in Cuy-Saint-Fiacre, France, participated in the experimental study, wearing collars equipped with IMU sensors for about three weeks. Recorded data were analyzed to measure the cats activity and to gain further insights into their biological rhythms.
We first determined the time budget of the cats by categorizing behavior into inactivity and activity. Next, we analyzed the day/night activity repartition and the hourly distribution of activity.
Results showed an average of 14.5% of global activity and a higher activity during the day in comparison with the night. Moreover, a bimodal activity pattern with increased activity at the time of the caretaker's interventions at feeding time was found.

## INTRODUCTION

Reinberg (2003) describes biological rhythm as "the periodical distribution of biological, physiological and behavioral phenomena". Biological rhythms are determinant for an organism's survival since it allows being in phase with different parameters of its environment such as temperature, photoperiod, humidity or food ressources (Sharma, 2003).
The research on the rhythmic behavior of domestic cats (Felis catus) is limited, and the findings show notable differences.
Cerutti et al (2018) suggest that these rhythms are driven by "circadian oscillators " and run on a 24–hour cycle in absence of any temporal information. When domestic cats are placed in free running conditions (environment devoid of zeitgebers), circadian rhythm of cerebral temperature (Kuwabara et al., 1986), body temperature (Randall et al. 1986) and activity (Randall et al., 1985; Randall et al. 1986; Cerutti et al., 2018) have been shown.

Similarly, an expression of a 24-hour activity rhythm has been revealed in pet cats (Piccione et al., 2013; Cove et al., 2017; Piccione et al., 2018; Parker, 2018), shelter cats (Baguet, 2012; Parker, 2018), farm cats (Panaman, 1980), and laboratory cats (Kuwabara et al., 1986; Randall et al. 1986).

Results are unclear when cats are placed under artificial day/night alternation. (Kavanau, 1971) reported no daily rhythmicity in a female cat placed in 12 hours day, 10 hours night, 1 hour of day/night transition and 1 hour of night/day transition. Results can be questioned as this study has only been conducted with one subject placed in a suboptimal environment. However similar results were shown in a less restrictive context and with a larger sample. For instance, (Piccione et al. , 2014) reveal an absence of daily locomotor rhythm in 5 housed cats with a free daily access to a garden.

These mixed results may be attributed to considerable differences in the methodologies used to observe cat behaviors and the varying experimental conditions.

The distinction between diurnal and nocturnal species is well established (Refinetti, 2008) and has been a key focus in research on activity rhythms. Once again, the findings are highly inconsistent, with some authors describing this species as nocturnal (Alterio & Moller 1977; Kuwabara et al., 1986; Haspel & Calhoon, 1993; Cove et al., 2017; Ferreira et al. 2020) and others as diurnal (Kavanau, 1971; Panaman, 1980; Piccione et al., 2014; Piccione et al., 2018). As suggested by Parker (2018), the absence of consensus may underline the inadequacy of this dichotomy diurnal/nocturnal to describe the cat. In line with this suggestion, Refinetti et al. (2016) and Parker et al. (2018) showed an important variability in the day/night activity's distribution among subjects. Moreover, repartition of daily activity differs according to environmental conditions (Horn et al., 2011) and seasons (Izawa, 1983). Rather than labeling cat's daily activity rhythm through chronotype (nocturnal vs diurnal), some authors found more relevant to characterize it as bimodal (dawn and dusk). Indeed, feral cats (Izawa, 1983; Konecny, 1987; Goszczyński et al., 2009), shelter cats (Baguet, 2012) and laboratory cats (Kuwabara et al., 1986; Piccione et al., 2018) have been observed to exhibit peaks of activity at dawn and dusk.

Furthermore, it has been demonstrated that factors such as reproductive status can influence the activity patterns of domestic cats. Ferreira et al. (2020) showed a decreased activity consecutive to male's castration. The authors of this study explain the higher activity of feral cats compared to pet cats (Horn et. al, 2011) by the difference in the reproductive status observed in these two populations, since only pet cats were castrated.

The present study aims to build on the existing knowledge of biological rhythm in domestic cats living in a group, using embedded accelerometers to monitor cats.

This technology has been used to record the activity of different species including domestic cats (Watanabe et al., 2005; Piccione et al., 2013; Piccione et al., 2014; Piccione et al., 2018) for several years now. It is considered as a cost-effective and a user-friendly tool for monitoring cats physical activity and has been used in multiple studies (Prigent Garcia and Chebly, 2024).

Recent studies demonstrate the utility of an accelerometer data in distinguishing an active from an inactive state with a very high accuracy (Yamazaki et al., 2020). In the present study, we aim to explore the activity pattern of a group of cats housed in the

AVA domain in France using accelerometer signals. Data from 12 different cats were recorded. We first established the time budget of the cats considering two categories: inactivity and activity. Then we examined the day/night activity repartition. Finally, the hourly distribution of activity was also investigated.

## MATERIALS AND METHODS

*Subjects and housing conditions*

The present study took place at AVA (Agir pour la Vie Animale) shelter located in the North of France (Cuy-Saint-Fiacre, Normandy).

Among 80 cats housed at AVA, 23 subjects likely accepting to wear a collar were pre-selected by the person in charge of the cattery. In order to test the acceptance of the device, these cats have been equipped with dummy collars approximately two weeks before the beginning of the effective data recording. Cameras were also installed then. The preliminary observations were conducted by the animals' regular caretakers and led to the selection of 18 cats tolerating the collar well. For those cats, the dummy collars were replaced by the real prototypes on September 21, 2020 and the activity of the cats was recorded continuously (24 hours a day) for 21 days from September 21, 2020 to October 13, 2020.
Six cats ended up not participating in the study for different reasons (loss of the collar, getting sick, collar's dysfunction). Thus, data of 12 cats: 8 females (Ethna, Mia, Gin, Aquarelle, Sigrid, Tara, Mimine and Lina) and 4 males (Monsieur S, Bouly, Ino, Ready), were analyzed. All the cats were neutered and had an average age of 9 years (min: 2 years, max: 16 years). The cats included in our study were pet cats that arrived at the shelter for various reasons (e.g., death of owner, illness…). Lina and Ino joined the shelter briefly before the beginning of the study and were just coming out of their adaptation phase during which they were restricted to one area of the cattery.

The whole domain of the shelter of 75 ha is accessible to the cats, however the feeding and the watering points as well as the sleeping areas reserved for the cats are limited to approximately 7000 m². This cat's reserved area, as shown in Fig. 1, consists of a natural area with grass, bushes, trees and 5 indoor areas: 4 cottages and 1 mobile home with free access. The latter contains litters, cat trees, water and food bowls, as well as beds arranged on shelves at different heights. Outside, the cats could take refuge in numerous huts containing straw. Feeding and watering points were also available. The cats were exposed to an average of 11:37 hours of light per day (min: 11:01; max: 12:13). On average, during the study, the sun rose at 7:55 am (min: 7:41 am; max: 8:11 am) and set at 7:33 pm (min: 7:12 pm; max: 7:55 pm) and the temperature varied from 9 to 16°C.

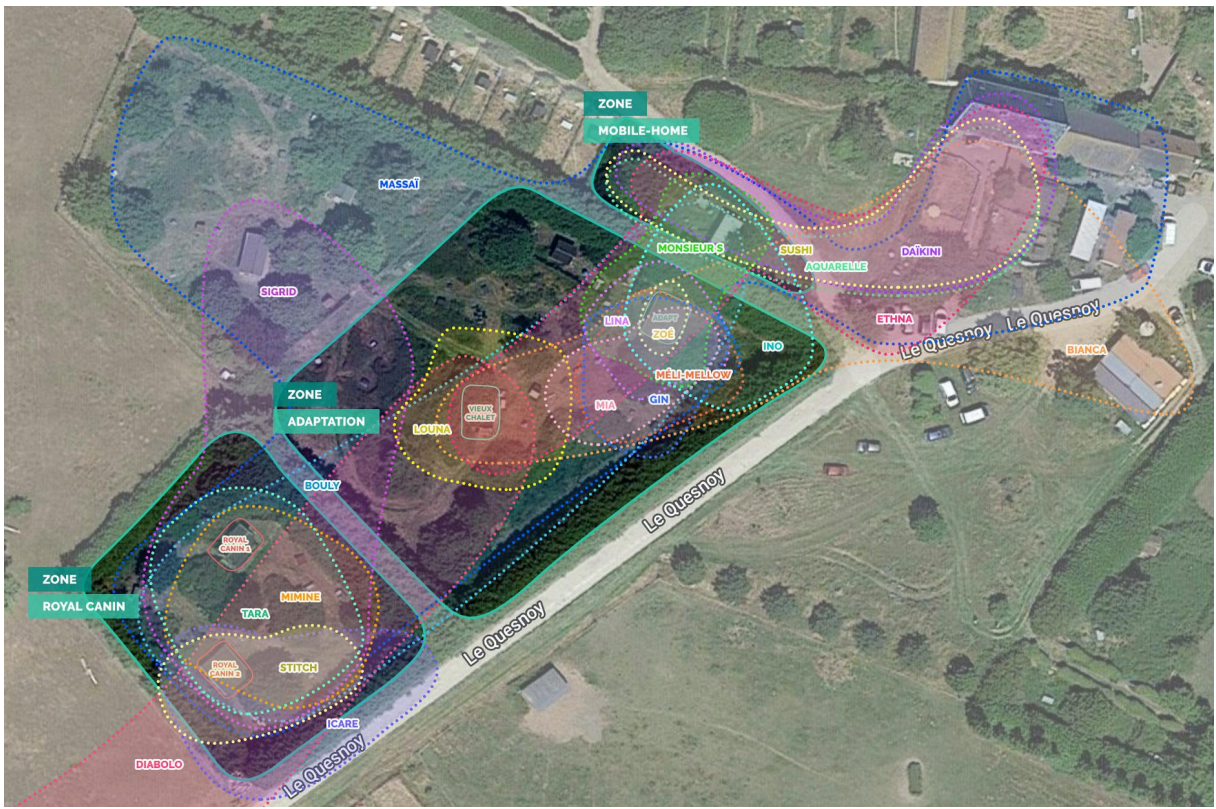

Fig. 1: Ava Shelter, zone reserved for cats.
Colored areas show the installed cameras field of view

The cats were fed with dry food as well as wet food. The water and food were renewed twice a day by the caretakers: once around 9:30 a.m. and a second time around 4:30 p.m. Treats could be given by the caretakers to facilitate certain care. The cats' environment was cleaned once a day at 9 a.m. In addition to these daily events, possible additional human interventions could occur in particular by the need to provide care to sick animals. There was however a certain regularity in the interventions of the caretakers. For example, as mentioned earlier, the cats were fed at relatively fixed times.

## Experimental Design

Data were recorded using an open source data logger called OpenLog Artemis by Sparkfun (Sparkfun), see Fig.2. This device permanently records at 0.2 Hz the on board temperature along with signals from an Inertial Measurement Unit (IMU). This sampling frequency is chosen to get the maximum battery autonomy. This 9 DoF IMU includes a 3-Axis gyroscope with +/- 500 dps range, a 3-Axis accelerometer with +/- 2 g range and a 3-Axis magnetometer with a +/- 4900 µT range. The system is powered with a 3.7 V/500 mA lithium battery. The logger and the battery are housed in a specifically 3D printed case placed on an adjustable collar with a total weight of 21 g. The collars, designed in nylon for the comfort of cats, were equipped with an anti-strangulation system. During the measurement campaign, collars' batteries were periodically recharged and then returned to the cats at 9 a.m. every 3 days.

In the present study, only the accelerometer among the sensors of the IMU was used to compute the cat's activity levels.

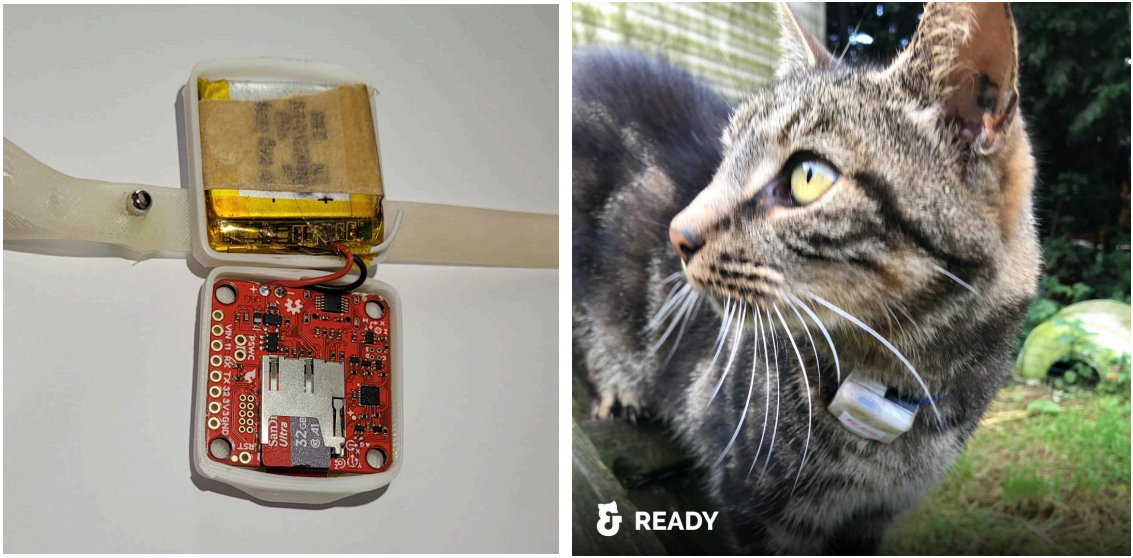

Fig. 2: (left) Sparkfun data logger in the 3D printed case, (right) Cat 'Ready' wearing the collar during the study

A total of 17 cameras (model RLK8-800B4-8MP) were placed at strategic locations in the cattery to cover areas in which cats were most often seen (Fig.1). Five cameras were placed indoors and 12 outdoors. All the cameras were equipped with infrared systems permitting the recording even at night.

## *Ethical Note*

This study was conducted in compliance with French animal welfare regulations as per the "Code Rural et de la Pêche Maritime article R214-87" and the European Directive 2010/63/EU.
The collars used in this study were lightweight collars (total weight of 21 g) equipped with an accelerometer, along with other sensors and a datalogger. The collars were designed in nylon for the comfort of cats and were equipped with an anti-strangulation system. This specific design allows the cats to move freely without causing any distress or discomfort.
Prior to the study, the cats were acclimated to wearing the collars to minimize any potential stress. The health and behavior of the cats were monitored closely throughout the study to ensure their well-being. Cats exhibiting stress during the acclimation phase (excessive grooming or scratching, reduced food intake, etc.) were discarded from the study.
No adverse effects were observed during the period in which the collars were worn.
At the end of the data collection period, the collars were removed, and the cats resumed their normal activities. All procedures were performed by trained staff throughout the study.

# DATA PROCESSING

## *Activity Levels Estimation*

Fig. 3.a shows the three dimensional acceleration measured by the SPARKFUN IMU at 0.2 Hz during 20 minutes of recording where the cat's behaviors are filmed. Fig. 3.b shows the acceleration intensity computed in the IMU frame as :

$$A = sqrt(Ax^2 + Ay^2 + Az^2),$$

where $Ax, Ay$ and $Az$ are the accelerations measured along the x, y and z axis respectively.

In order to classify the cat's behaviors between activity and inactivity, we compute the absolute value of the acceleration derivative over time, dA, see Fig. 3.c, as :

$$dA = abs(\frac{An+1 - An}{tn+1 - tn}), \quad n = 0...N$$

where $N$ is the total number of the recorded measures, and t is the time vector.
Note that the measured acceleration represents the cat's movements as well as the gravity measure. A mean acceleration of approximately 10 m/s² can be seen in Fig. 3.b. Given that the gravitational acceleration is constant over time, derivating the acceleration signal cancels its effect, yielding only information related to the cat's movements (Fig. 3.c).
As a result, using an optimal threshold on the absolute value of the acceleration derivative permits distinguishing data corresponding to an activity or an inactivity.
Indeed, as shown in Fig. 3.d, every 5 seconds, we evaluate a measure as activity (activity level=1) if the computed acceleration derivative exceeds the threshold, otherwise the measure is evaluated as inactivity (activity level=0).

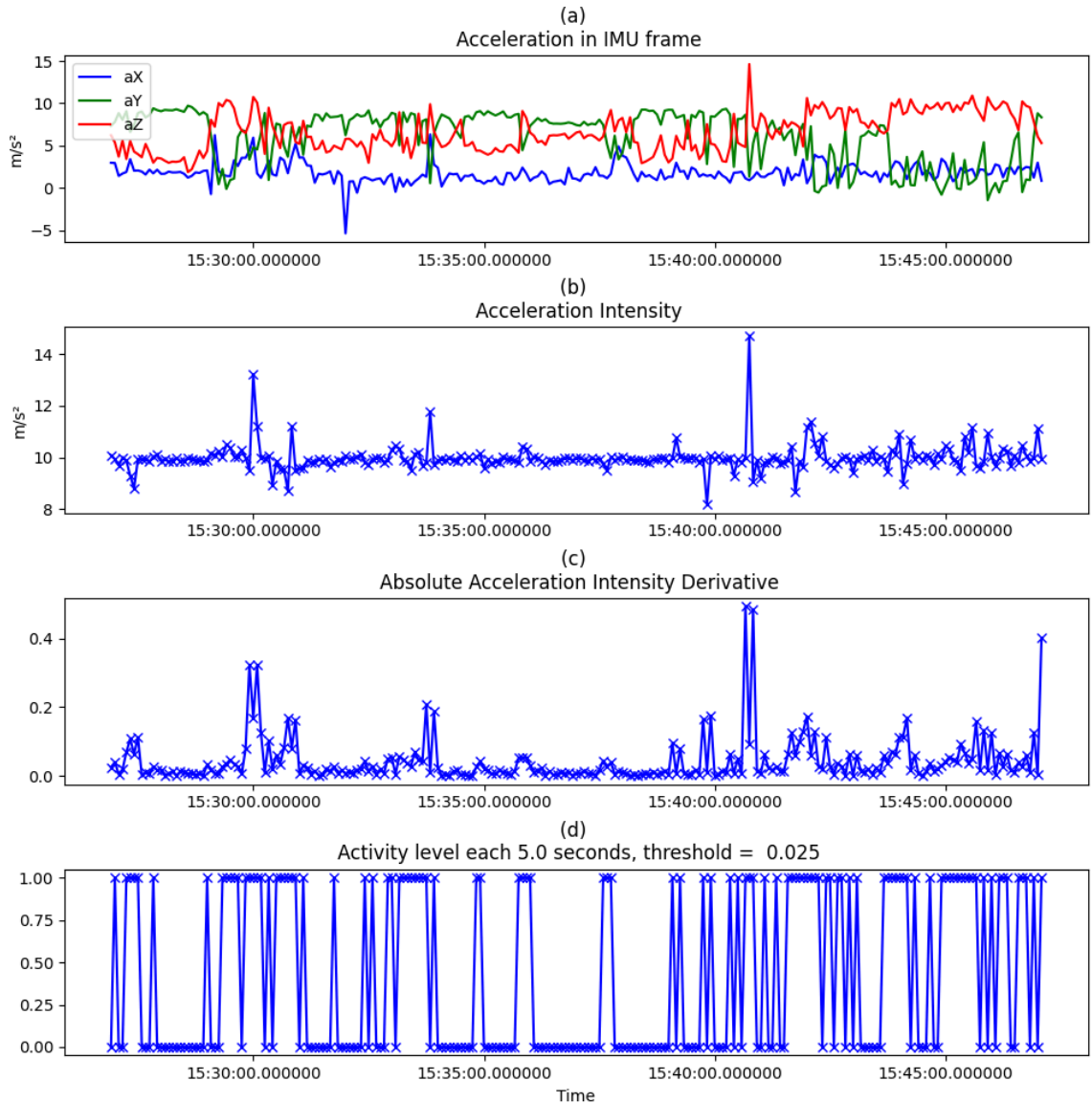

Fig. 3: Activity level computation: (a) 3-axis acceleration, (b) acceleration intensity, (c) absolute acceleration derivative, (d) computed activity level

*Activity Levels Expectations*

Using the recorded videos, we label the cat's behaviors over time. In this study, resting or sleeping are considered as inactivity (expectation=0), while walking, grooming, eating, drinking and jumping are considered as activities (expectation=1).

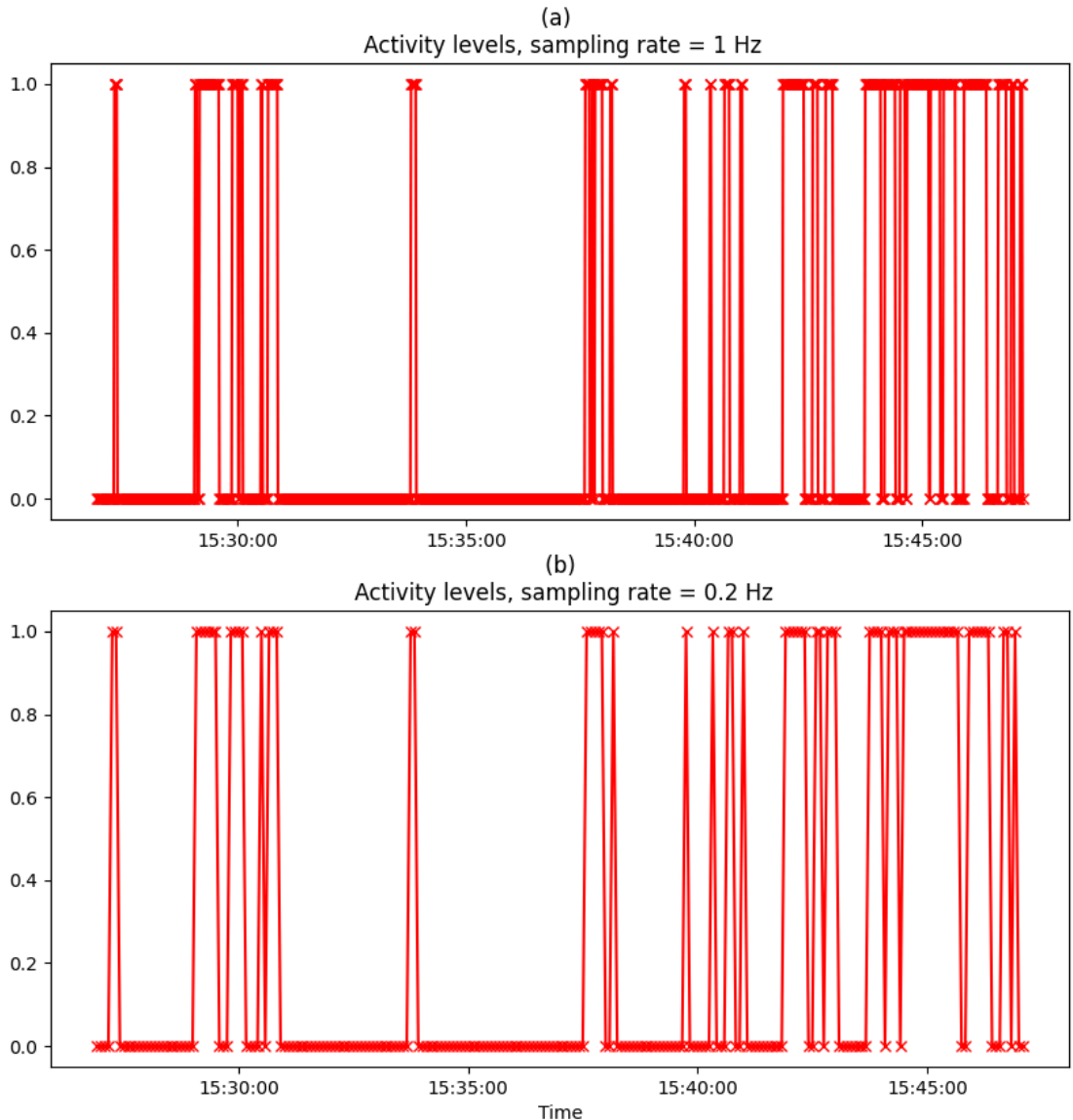

Fig. 4: Activity level expectations: (a) at 1 Hz, (b) at 0.2 Hz

The cat's behavior is labeled every second, yielding an expectation of 0 or 1 with a sampling rate of 1 Hz, as can be seen in Fig. 4.a.

Since the estimation using the sensor's data is carried out at 0.2 Hz, which is equivalent to an estimation every 5 seconds, we need to compute an expectation every 5 seconds as well, in order to compare the estimations and the expectations.

Then, using a window of 5 seconds, we compute the total time of activity within this window. If the cat was active for at least 20% of the total window size, i.e. 1 second within a 5 seconds window, the activity level is set to 1. Otherwise the cat is considered resting and the activity level is set to 0. The final result is presented in Fig. 4.b where an activity level expectation is computed at a 0.2 Hz sampling rate.

*Algorithm Validation*

The optimal threshold value for the physical activity to distinguish whether cats were resting or not is chosen using a ROC curve. The performance of the classification model is evaluated at different classification thresholds, going from 0 to 1. Fig. 5 shows the computed ROC curve, where the x-axis represents the classifier's specificity while the y-axis represents the classifier's sensitivity. The selected threshold is 0.025 with a sensitivity of 73,4%, a specificity of 29,5% and an AUC of 0.74.

Fig. 5: ROC curve

Fig. 6 shows the estimated activity level using the optimal threshold of 0.025 in blue lines compared to the expectations in red lines, every 5 seconds. The activity intensity is defined as the amount of activity and is computed as the count of activity every 60 seconds. Fig. 7 shows the estimated activity intensity in blue lines compared to the expected activity intensity in red lines. There is significant positive correlation between the computed activity intensity and the expected one, with a pearson coefficient of 0.71 and a p-value of 0.0004.

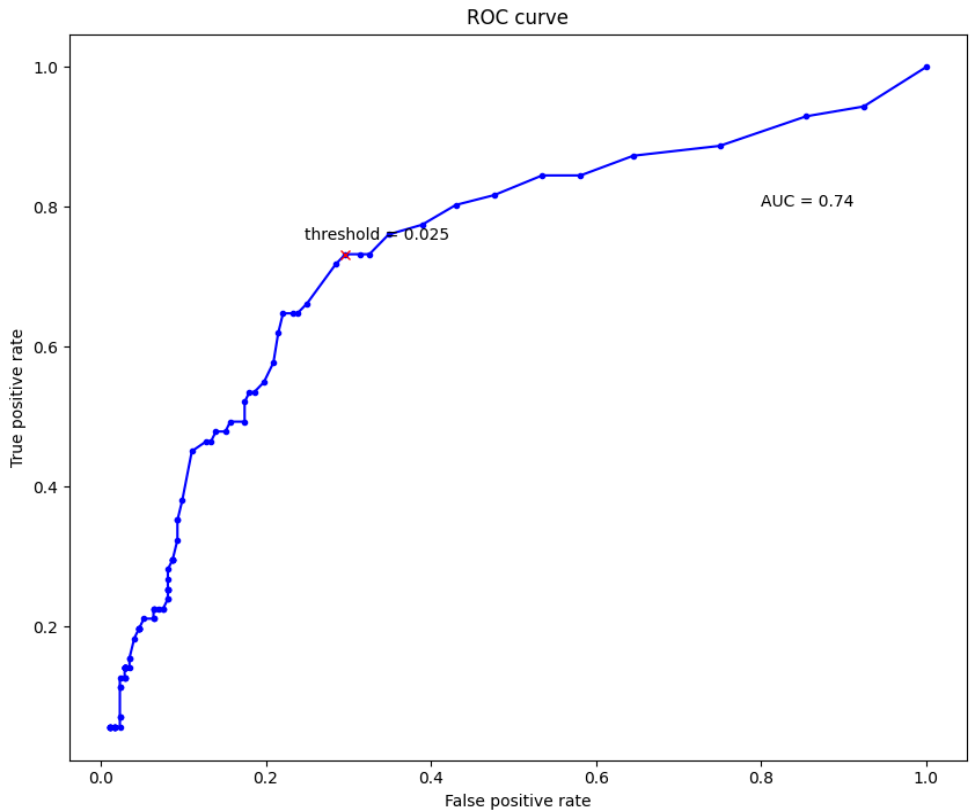

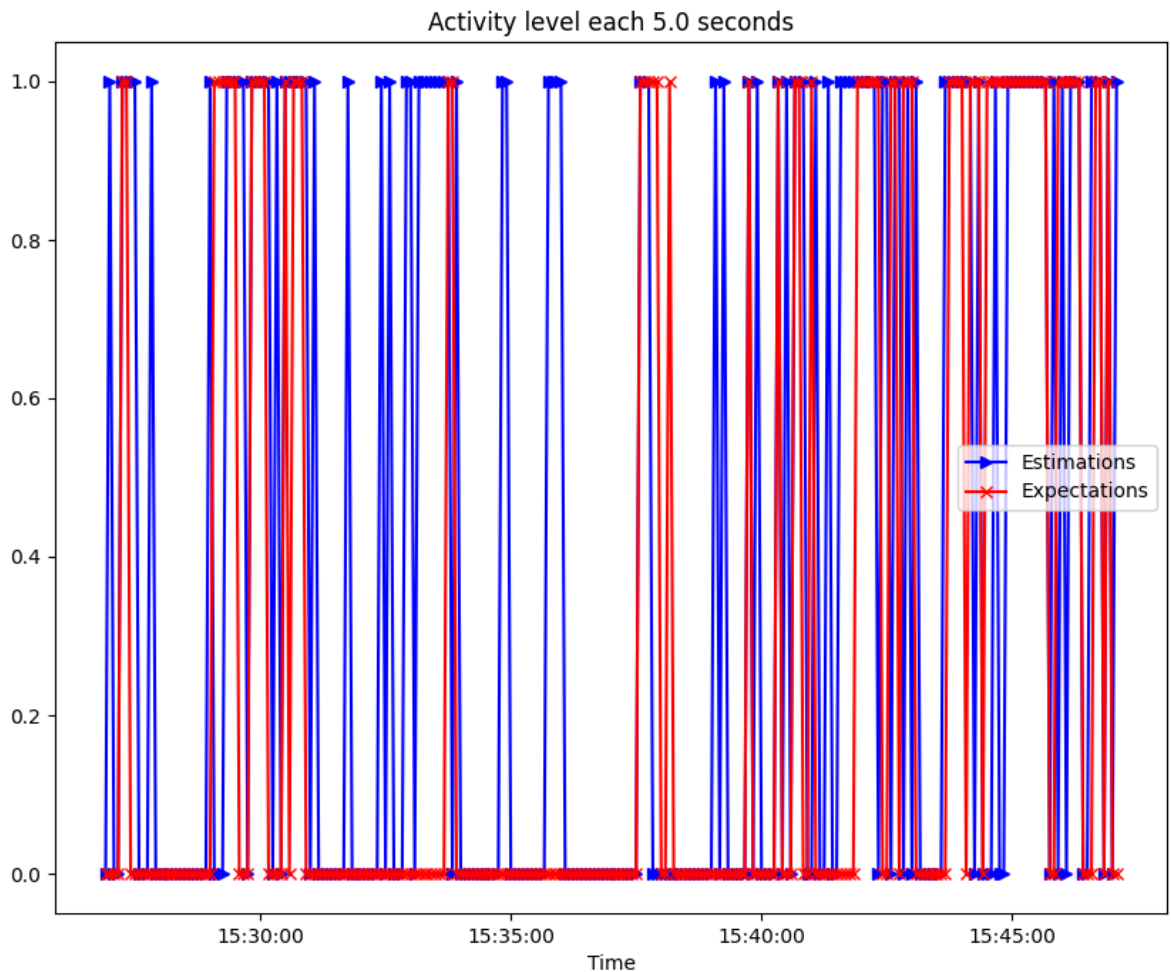

Fig. 6: Estimated vs. Expected activity levels

*Statistical Analysis*

All analyses were performed with version 4.0.4 of R (R Development Core Team, 2020). Due to the non-normal distribution of the data, a Wilcoxon test was performed to compare the proportion of time spent in activity versus inactivity as well as the proportion of time spent in activity during the photophase versus the scotophase. Finally, a 1-way ANOVA was conducted to compare the proportion of time spent in activity according to the time of day. Two-to-two comparisons were made with Tukey's post-hoc tests.

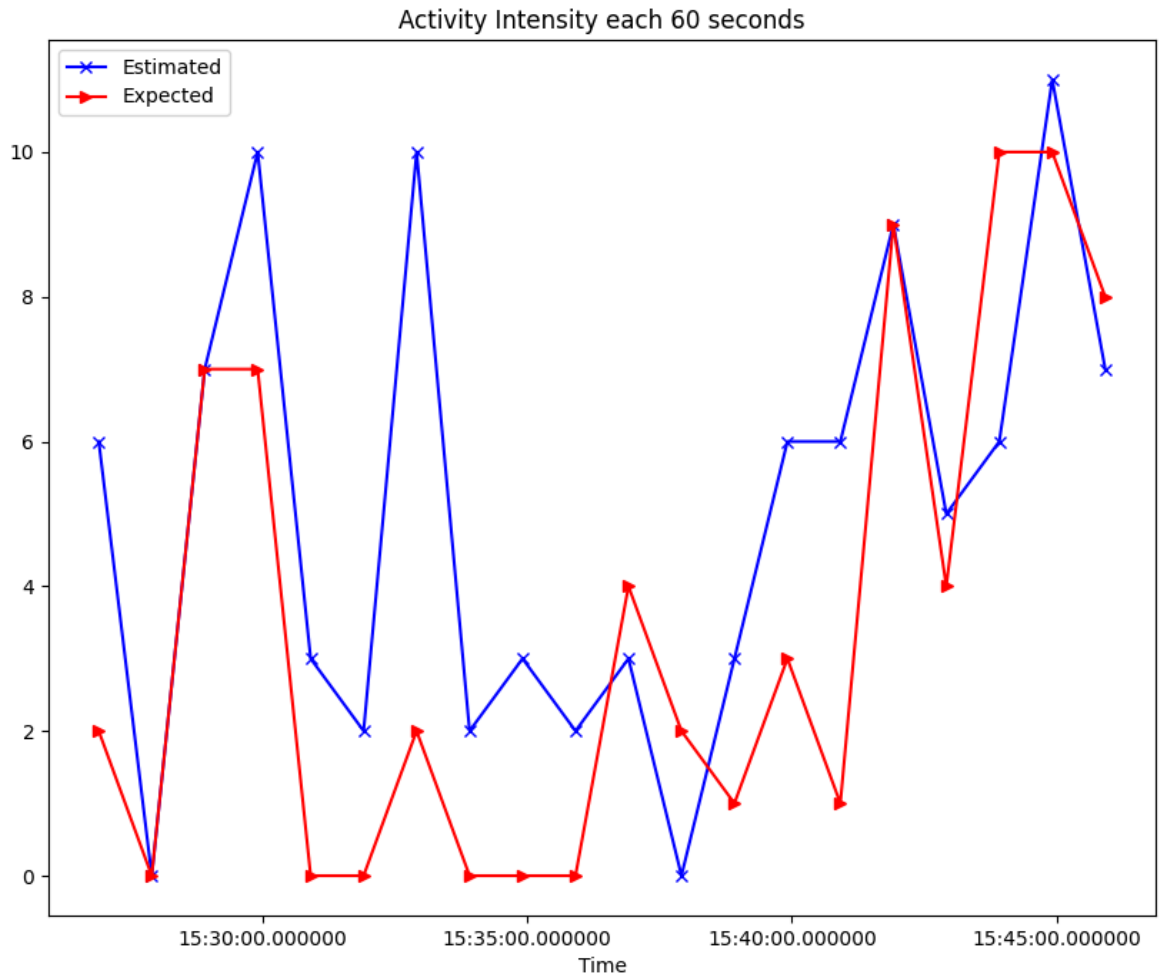

Fig. 7: Computed activity intensity compared to expected activity intensity

## RESULTS

### *Time Budget*

A Wilcoxon test confirms the significant difference between the activity and the inactivity distributions, with w = 78 and p = 3.6e-5 at 5% significance level.
On average, as shown in Fig. 8, cats spent 85.5% of their time inactive and 14.5% active, with a standard deviation of 3.26.
Only one cat, Lina, presented an activity percentage lower than the mean activity by more than twice the standard deviation (7.7% of activity).

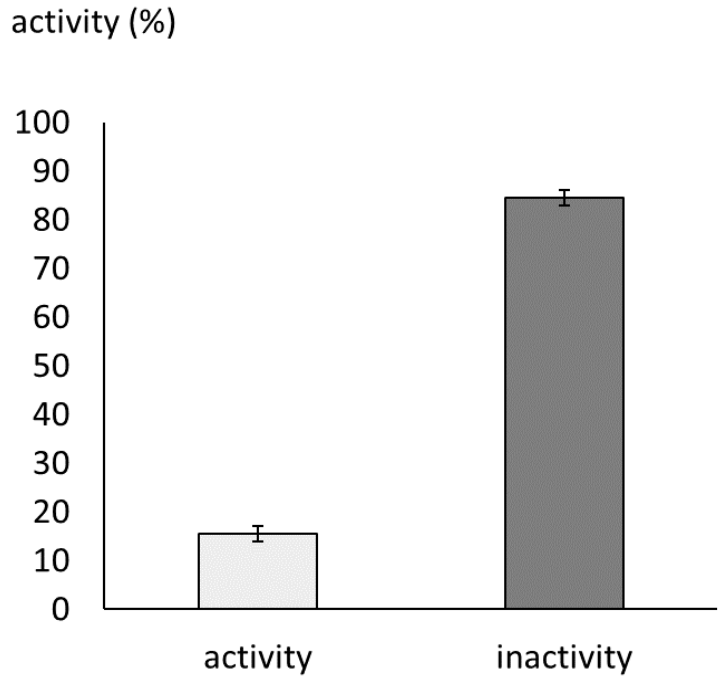

Fig. 8: Average proportion of time spent in activity versus inactivity in 24 hours. Data was collected from 12 cats over 3 weeks. Error bars represent standard deviations.

## Chronotype

Fig. 9 shows the percentage of time spent in activity during the day (mean = 64.86%, std = 6.61) and during the night (mean = 35.13%, std = 6.61). On average, cats were significantly more active during photophase versus scotophase (Wilcoxon test: W=222, p = 3.63e-5).

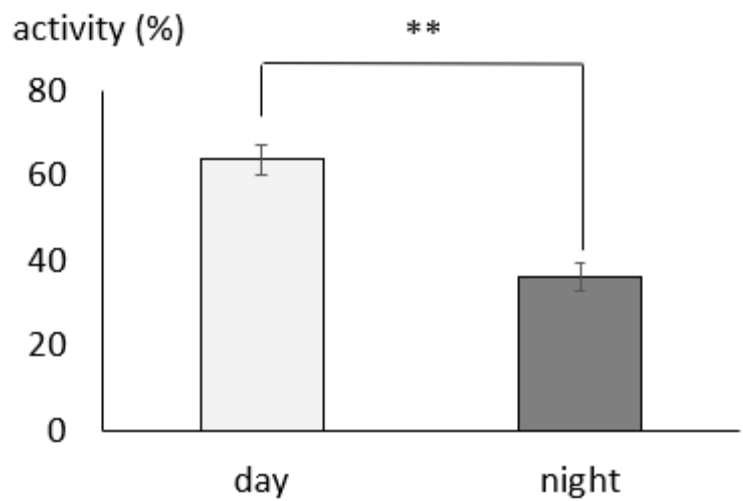

Fig. 9: Average proportion of time spent in activity during the day versus the night. Data was collected from 12 cats over 3 weeks. The ** symbol represents a p value <0.01 in Wilcoxon test. Error bars represent standard deviations.

## Hourly Activity

Fig. 10 shows the distribution of cats activity over a 24 h period. The average proportion of time spent in activity varied significantly throughout the day (One-way ANOVA, F=10.15, p = 4.4e-25<0.05). Cats were significantly more active between 9:00 am and 12:00 am and between 4:00 pm and 6:00 pm.

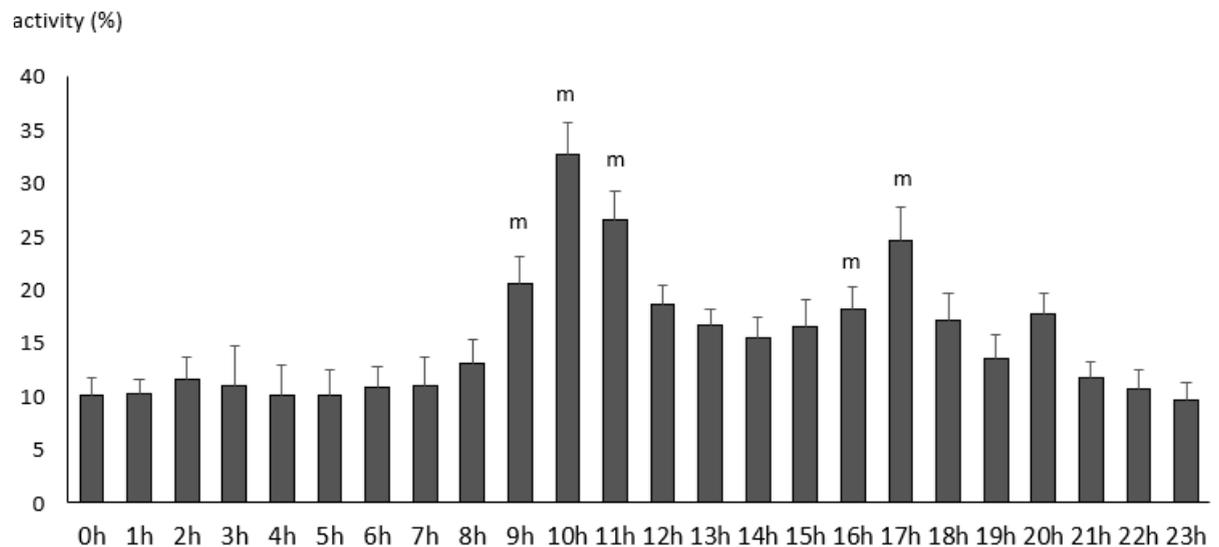

Fig. 10: Average proportion of time spent in activity each hour. Data was collected from 12 cats over a 3-week period. The "m" symbol indicates a p-value inferior to 0.05 on Tukey's test. Error bars represent standard deviations

DISCUSSION

*Time Budget*

The level of activity computed for the participating cats (14,5%) is close to the one shown by Horn et al. (2011) in pet cats. In contrast, greater activity has been found in feral cats (Horn et al., 2011), with a daily activity level of 38%. Increased activity in feral cats could be explained by the necessity to hunt for food. As for AVA cats, even though they have access to a free outdoor space, it seems that they are not interested in hunting activities. Indeed, no hunting behaviors were observed during the experiments or reported by the caretakers. The activity level of AVA cats is then partly related to their nutritional lifestyle.

The influence of other factors was considered in the literature. Particularly, Ferreira et al. (2020) suggested that the difference in activity between feral and pet cats reported by Horn et al. (2011) was the result of a difference in the reproductive status between these two populations. Indeed, all pet cats were neutered while only 9% of feral cats were neutered in the study conducted by Horn et al. (2011). The hypothesis that sterilization of pet cats would have caused a decrease in their activity is congruent with the results of the present study as all the subjects participating to the study are neutered.

As an exception, a low activity level (7.7%) was found for Lina, a 5-year-old cat barely finishing her adaptation period at the shelter when the study began. Both during the meeting with the cats and through direct observation, we could note that

Lina seemed fearful and spent a lot of time hidden in the cat trees of one of the cottages. When the caretakers arrived to renew the food, most of the cats followed them while Lina stayed hidden and only came out when the other cats and the caretakers were gone. Thus we can assume that this cat's activity was inhibited by some fear. As for Ino, the 2-year-old cat who was also finishing its adaptation phase when the study began, it seems that the adaptation phase didn't affect its activity much (13.6% of activity), probably because it is a younger cat and then it is able to adapt relatively quickly to new environments.

## *Chronotype*

The notion of chronotype has sometimes been questioned because of the contradictory results reported in the literature. Indeed, while several studies have highlighted greater activity during the day (Kavanau, 1971; Panaman, 1980; Horn et al., 2011; Piccione et al., 2014; Parker, 2018; Piccione et al., 2018), others have shown the opposite (Alterio & Moller, 1977; Kuwabara et al., 1986; Haspel & Calhoon, 1993; Horn et al., 2011; Cove et al., 2017; Ferreira et al., 2020).

It seems relevant to note that the majority of the studies showing superiority of nocturnal activity over diurnal activity have been conducted on feral cats (Alterio & Moller, 1977; Haspel & Calhoon, 1993; Horn et al., 2011; Cove et al., 2017). Conversely, the majority of studies demonstrating a diurnal activity included populations of human-dependent cats, i.e., pet cats (Horn et al., 2011, Piccione et al., 2014), shelter cats (Parker, 2018), or laboratory cats (Kavanau, 1971; Piccione et al., 2018). The hypothesis that feral cats seek to avoid human presence has been proposed to explain the nocturnal activity of feral cats (Haspel & Calhoon, 1993; Horn et al., 2011; Cove et al., 2017). This nocturnal activity could also reflect the adjustment of feral cats to the activity patterns of their prey (Horn et al., 2011). Indeed, a meta-analysis of 27 studies from 4 continents (North America, Europe, Africa, and Australia) showed that the diet of feral cats consists of up to 76% of small mammals, including rats, rabbits, and mice, which are nocturnal species (Plantinga et al., 2011). The diurnal activity reported in cats fed and cared for by humans could then be explained by the fact that these animals are not subject to the need to hunt to meet their energy needs. Therefore, rather than adjusting to the rhythm of their prey, they would adjust to the rhythm of humans. This was suggested by some authors who noted that cats were more active when their owners (Piccione et al., 2013) or laboratory staff (Randall et al., 1985) were present. We can then claim that the results obtained in this study are in phase with this suggestion, since the cats in our study were significantly more active during photophase.

In other words, the day/night distribution of a cat's activity seems to be partly dependent on its lifestyle (dependence or independence of humans). Therefore, trying to characterize this species by the notion of chronotype does not seem relevant. It would be more accurate to highlight the strong adaptability of the domestic cat to its environmental conditions. This is what (Izawa, 1983) had suggested by showing that feral cats were more active during the day in winter, and at night in summer.

## *Hourly Activity*

The influence of the human presence on the cat's behavior is studied by the analysis of the activity per hour. Indeed, two peaks of activity during the day were concomitant to the interventions of the keepers. The first peak was observed between 9:00 am and 12:00 am, covering the period before, during and after the first visit of the caretakers taking place around 9:30 am. The second peak was observed between 16:00 pm and 18:00 pm, also lying before, during and after the second intervention of the caretakers taking place around 16:30 pm.

Several authors have demonstrated a bimodal activity pattern with increased activity at sunrise and sunset (Izawa, 1983; Kuwabara et al., 1986; Konecny, 1987; Goszczyński et al., 2009; Baguet, 2012; Piccione et al., 2018; Parker et al., 2019). This bimodal pattern is explained by the fact that favorable conditions for predatory activity, such as moderate temperature and humidity, and good availability of prey, are more present at sunrise and at sunset (Randall et al., 1986).

While a bimodal activity pattern is indeed found in AVA shelter cats, it appears to be punctuated more by human activity than by photoperiodic variations.

Similarly, Randall et al. (1986) had shown a spike in activity associated with human intervention in a group of laboratory cats and Parker et al. (2019) had demonstrated a greater activity at the time of the handler 's intervention in a cattery.

These results could be explained by the fact that, fed by humans, the cats in a laboratory, a cattery or a are not forced to focus their activity on dawn and dusk, since they are not interested in hunting activities.

On the other hand, two studies have found peaks of activity at dusk in laboratory cats (Kuwabara et al. 1986; Piccione et al. 2018) even though these cats don't need or can't perform hunting. However, these studies were performed in standardized environments, where everything was organized to reduce the impact of human presence on the animals' behavior. The subjects in both studies were placed in deafened rooms. The cats observed by Kuwabara et al. (1986) were placed in opaque cages, and the staff only intervened to clean the cages when the cats were already active.

Thus, the authors suggested that the bimodal activity pattern with a dawn peak and a dusk peak would have been selected in the ancestor of the domestic cat (*Felis libyca*) and would have persisted in the domestic cat, despite a food supply.

The results presented in this study suggest that this bimodal pattern is not fixed and can be modified by human's presence.

Note that the percentage of hourly activity remains above a 10% floor, and that could be explained by the variation of activity for each cat among different days and by the variations between cats. Indeed, it appears that every cat has its own activity pattern during the day, and its rest hours may vary from one day to another. Thus, the average of activities among the twelve cats and for three weeks shows a non-null hourly activity over the 24 hours of the day.

# CONCLUSION

In this study, biological rhythms in domestic cats are discussed. An experimental study has been conducted at AVA domain in Cuy-Saint-Fiacre, France, with twelve participating cats. These cats were equipped with a data logger collar for approximately three weeks. The recorded acceleration enabled a quantification of the cat's activity and a thorough study of their biological rhythms.

The time budget of the cats is studied and results revealed that a domestic cat spends nearly 14.5% of its time in activity. In addition, participating cats were more active during the day and their activity was maximal during the interventions of the caretakers twice a day. A bimodal activity pattern that seems to be synchronized with human activity rather than with photoperiodic variations is then shown.

As a perspective, we aim to conduct more experiments with a larger number of participating cats in order to assess the effect of age and sex on the activity level. Observations could also be conducted at different times of the year to test the effect of seasonal conditions on cats activity patterns as suggested in some studies.

# ACKNOWLEDGMENT

The authors would like to extend their sincere gratitude to Thierry Bedossa, director of the AVA shelter, for his instrumental role in making this study possible. We also wish to thank the caregivers at AVA for their dedication, assistance, and cooperation throughout the study. Additionally, we would like to express our appreciation to our colleagues at Blackfoot for their expert technical support, which was critical to the success of this research.